\documentclass[aps,preprint,preprintnumbers,amsmath,amssymb]{revtex4}
\usepackage{amsmath,mathrsfs,amsbsy,color,graphicx,bm,amsthm,amsfonts}
\usepackage{units}
\usepackage{bbm}
\usepackage{times}
\usepackage{dcolumn}
\usepackage{mathrsfs}
\usepackage{amsmath,amssymb,epsfig}
\usepackage{hyperref}

\begin{document}

\title{Generation of quantum coherence for continuous variables between causally disconnected regions in  dilaton spacetime}
\author{Qinglong Xiao$^{1}$}\thanks{These authors contributed equally to this work.} \author{Cuihong Wen$^{1}$}\thanks{These authors contributed equally to this work.}   \author{Jiliang Jing$^{1}$}  \author{Jieci Wang$^{1}$\footnote{Email: jcwang@hunnu.edu.cn}}
\affiliation{$^1$  Department of Physics and Synergetic Innovation Center for Quantum Effects,\\Key Laboratory of Low-Dimensional Quantum Structures and Quantum Control of Ministry of Education, \\
Key Laboratory for Matter Microstructure and Function of Hunan Province,\\
 Hunan Normal University, Changsha 410081, China}


\begin{abstract}
We study the dynamics of Gaussian quantum coherence under the background of  a Garfinkle-Horowitz-Strominger dilaton black hole. It is shown that the dilaton field has evident effects on the degree of coherence for all the bipartite subsystems.  The bipartite Gaussian coherence is not affected by the frequency of the scalar field for an uncharged or an extreme dilaton black hole. It is found that the initial  coherence is not completely destroyed even for an extreme dilaton black hole, which is  quite different from the behavior of quantum steering  because the latter suffers from a``sudden death" under the same conditions.  This is nontrivial because one can employ quantum coherence as a resource for quantum information processing tasks even if quantum correlations have been destroyed by the strong gravitational field.  In addition, it is demonstrated that the generation of quantum coherence between the initial separable modes  is easier for low-frequency scalar fields.  At the same time, quantum coherence is smoothly generated betweenone pair of partners and   exhibits a ``sudden birth" behavior between another pairs in the curved spacetime.

\end{abstract}

\vspace*{0.5cm}

\maketitle
\section{Introduction}

String theory is regarded as one of the most promising candidates for a consistent understanding of quantum mechanics and the theory of gravity.
Different from general relativity,  it is predicted in string theory that  the presence of dilaton fields can change the properties of black holes \cite{chen18, gar7,gar8,gibbs}.
A solution of  static spacetime, i.e.,  the Garfinkle-Horowitz-Strominger (GHS) dilaton black hole \cite{chen18, gar7}, can be obtained by choosing the invariant to be the Lagrangian of the electromagnetic field. The Hawking temperature \cite{Hawking1} of the GHS black hole not only depends on its mass, but also on the dilaton field because the latter is also a resource of gravity.
On the other hand, the behavior of quantum information near the event horizon of black holes has received considerable attention  in recent years  \cite{Schuller-Mann, jiliang, Ralph,RQI3, RQI4,RQI5, RQI6, RQI7, RQI8, jieci2,RQI9,biwei}.  It is believed that related studies can contribute to  a deeper understanding of  nonlocality between causally disconnected spacetime regions, as well as to a further understanding of the entropy and the problem of information loss for black holes \cite{har1,un3,hawking5}.

Quantum coherence \cite{C,coherencer,coherencer1,QS1}, as a key aspect of quantum physics, is not only an embodiment of the superposition principle of states but also the basis of many unique phenomena in quantum physics, such as quantum entanglement and steering. With the help of quantum coherence, one can implement various quantum information processing tasks which cannot be accomplished classically, such as quantum computing  \cite{computing2,computing4},  quantum  information storage \cite{trans2, trans3} , quantum communication \cite{trans1, QX}, and  quantum metrology \cite{metrology1, metrology2, metrology3}.  Therefore, understanding  quantum coherence in a more general framework  is important  both for  the  fundamental research of physics and the development of modern quantum technology. Despite its fundamental role in physics, the study of coherence receives increasing attention until Baumgratz \emph{et al.} introduced an architecture for the measurement of coherence  \cite{C}. In 2016,  a quantification of coherence for continuous variables was proposed by Xu \cite{XU}, which provides a framework of  quantum resource theory for quantum superposition in infinite-dimensional quantum systems \cite{inform1}.

In this paper, we study the distribution and generation of continuous-variable coherence  in the background of a GHS dilaton black hole. The considered initial state is a two-mode squeezed Gaussian state, which can be employed  to  define quantum vacuum when the spacetime has at least two asymptotically flat regions. This state has a special role in the quantum field theory because the values of squeezing parameter between causally disconnected regions are changed according to the spacetime structure.  We find that Gaussian quantum coherence is more robust than the steering-type quantum correlations under the influence of gravity in the GHS spacetime.

The structure of the paper is as follows. In Sec. II, we discuss the  dynamics and second quantization of the scalar field near the GHS dilaton black hole. In Sec. III and IV, we discuss the method of measuring the  coherence of continuous variables and the  behavior of Gaussian quantum coherence in the GHS spacetime, respectively. In the final section, we make a brief summary. Throughout the paper, the units $G$ = $c$ = $\hbar$ = $\kappa_B$ = 1 are used.

\section{VACUUM STRUCTURE OF THE SCALAR FIELD  \label{model}}

In this section, we review the quantum field theory in the GHS  black hole spacetime. The equations describing gravity in the context of string theory  can be approximated by Einstein's equations in the regions near the horizon of a black hole. In this scenario, the Schwarzschild solution is a good approximation to describe a static and uncharged black hole within string theory. However, when we consider the solutions of the Einstein-Maxwell equations, the presence of a scalar field called dilaton should be considered. In this case, the  added dilaton field  couples with the Maxwell field, which  changes the spacetime characters of the black hole.  In the low energy limit of string theory, the static dilaton black hole solution was provided by  Garfinkle, Horowitz, and Strominger \cite{gar7}. The dilaton gravity  in string theory is important because it is  a good candidate for an eventual quantum theory of gravity  \cite{gar8,gibbs}. Therefore, it is  meaningful to study the behavior of Gaussian quantum coherence in the dilaton spacetime.

The line element for a GHS dilaton black hole can be written  as \cite{gar7}
\begin{eqnarray}
ds^2&=&-\left(\frac{r-2M}{r-2D}\right)dt^2+\left(\frac{r-2M}{r-2D}\right)^{-1}dr^2+r\left(r-2D\right)d\Omega^2,
\end{eqnarray}
where $d\Omega^2=d\theta^2+\sin^2\theta d\varphi^2$,  $M$ is the mass of the black hole and $D$ is the dilaton charge. The event horizon of the GHS black hole is located at $r_+=2M$. 

The dynamics of  massless scalar field in a general background  is given by the Klein-Gordon equation \cite{birelli}
\begin{eqnarray}\label{inkg}
\frac{1}{\sqrt{-g}}\partial_{\mu}(\sqrt{-g}g^{\mu\nu}\partial_{\nu})\phi=0.
\end{eqnarray}

 Solving the  Klein-Gordon equation near the event horizon of the black hole, one obtains  the outgoing modes
\begin{eqnarray}\label{inac}
\phi_{out,\omega lm}(r<r_{+})=e^{i\omega u}Y_{lm}(\theta,\varphi),
\end{eqnarray}
\begin{eqnarray}\label{inad}
\phi_{out,\omega lm}(r>r_{+})=e^{-i\omega u}Y_{lm}(\theta,\varphi),
\end{eqnarray}
where $v=t+r^*$, $u=t-r^*$, and $r^*$ is the tortoise coordinates.

Employing the Schwarzschild modes given in Eqs. (\ref{inac}) and (\ref{inad}), the scalar field $\Phi$  near the event horizon can be expanded as \cite{birelli}
\begin{eqnarray}\label{First expand}
&&\Phi=\sum_{lm}\int d\omega[b_{in,\omega lm}\phi_{in,\omega
lm}(r<r_{+})+b^{\dag}_{in,\omega lm}\phi^{*}_{in,\omega
lm}(r<r_{+})\nonumber\\
&& \quad \quad \quad  \quad \quad \quad +b_{out,\omega
lm}\phi_{out,\omega lm}(r>r_{+})+b^{\dag}_{out,\omega
lm}\phi^{*}_{out,\omega lm}(r>r_{+})],
\end{eqnarray}
where $b_{in,\omega lm}$ and $b^{\dag}_{in,\omega lm}$ are the
annihilation and creation operators acting on the states of the
interior region of the dilaton black hole. Similarly, $b_{out,\omega lm}$ and
$b^{\dag}_{out,\omega lm}$ are the
operators acting on the vacuum of the exterior region, respectively.
The Schwarzschild vacuum state for the scalar field  can be defined as \cite{birelli}
\begin{eqnarray}\label{dilaton vacuum}
b_{in,\omega lm}|0\rangle_{in}=b_{out,\omega lm}|0\rangle_{out}=0.
\end{eqnarray}

On the other hand, the  light-like Kruskal coordinates $U$ and $V$ are defined by \cite{D-R, wang19},
\begin{eqnarray}
&&U=-4(M-D)e^{-u/(4M-4D)},\quad \nonumber\\
&&V=4(M-D)e^{v/(4M-4D)},\quad {\rm if\quad r>r_{+}};\nonumber\\
&&U=4(M-D)e^{-u/(4M-4D)},\quad\nonumber\\
&&V=4(M-D)e^{v/(4M-4D)}, \quad {\rm if\quad r<r_{+}}.
\end{eqnarray}
Then we can rewrite the field modes to
\begin{eqnarray}\label{inside mode1}
\phi_{out,\omega
lm}(r<r_{+})=e^{-4(M-D)i\omega\ln[-U/(4M-4D)]}Y_{lm}(\theta,\varphi),
\end{eqnarray}
\begin{eqnarray}\label{outside mode1}
\phi_{out,\omega
lm}(r>r_{+})=e^{4(M-D)i\omega\ln[U/(4M-4D)]}Y_{lm}(\theta,\varphi).
\end{eqnarray}

Making an analytic continuation for Eqs. (\ref{inside mode1}) and (\ref{outside mode1}), we obtain  \cite{D-R}
\begin{eqnarray}
\phi_{\uppercase\expandafter{\romannumeral1},\omega lm}=e^{2\pi \omega (M-D)}\phi_{out,\omega lm}(r>r_+)+e^{-2\pi \omega (M-D)}\phi^*_{out,\omega lm}(r<r_+),
\end{eqnarray}
\begin{eqnarray}
\phi_{\uppercase\expandafter{\romannumeral2},\omega lm}=e^{-2\pi \omega (M-D)}\phi^*_{out,\omega lm}(r>r_+)+e^{2\pi \omega (M-D)}\phi_{out,\omega lm}(r<r_+),
\end{eqnarray}
which shows that one can use the  Kruskal coordinates to introduce new orthogonal basis for the scalar field.

By expanding the scalar field $\Phi$ in terms of $\phi_{\uppercase\expandafter{\romannumeral1},\omega lm}$ and $\phi_{\uppercase\expandafter{\romannumeral2},\omega lm}$ in the GHS spacetime, one obtains
\begin{eqnarray}\label{second expand}
\Phi=\sum_{lm}\int d\omega[a_{I,\omega lm}\phi_{I,\omega
lm}+a^{\dag}_{I,\omega lm}\phi^{*}_{I,\omega
lm}+a_{II,\omega
lm}\phi_{II,\omega lm}+a^{\dag}_{II,\omega
lm}\phi^{*}_{II,\omega lm}],
\end{eqnarray}
  where the annihilation operator $a_{I, \omega lm}$ can be used to
define the Kruskal vacuum outside the event horizon  \cite{D-R}
\begin{eqnarray}
a_{I,\omega lm}\left|0\right \rangle_K=0.
\end{eqnarray}

It is seen that  Eq. (\ref{First expand}) is the expansion of the scalar field in Schwarzschild modes, while Eq. (\ref{second expand}) corresponds to the decomposition of the field in Kruskal modes \cite{D-R, wang19}. Then we can calculate the Bogoliubov transformations between  the particle annihilation and
creation operators  which act on the Schwarzschild vacuum  and  Kruskal vacuum, respectively. After some calculation, the
Bogoliubov transformations are found to be \cite{wang19}
\begin{eqnarray}
&&a_{I,\omega lm}=b_{out,\omega
lm} \cosh u -b^{\dag}_{in,\omega lm} \sinh u , \nonumber\\
&&a^{\dag}_{I,\omega lm}=b^{\dag}_{out,\omega
lm}\cosh u -b_{in,\omega
lm}\sinh u ,
\end{eqnarray}
where $\cosh u=\frac{1}{\sqrt{1-e^{-8\pi \omega (M-D)}}}$,  $a_{I,\omega lm}$ and  $a^{\dag}_{I,\omega lm}$ are the
annihilation and creation operators acting on the Kruskal vacuum of
the exterior region, respectively.  For the observer outside the black hole,  the modes in the interior region  should be
traced over because a local  observer has no
access to the information in the causally disconnected region. 

After normalizing the state vector, it is found that  the Kruskal vacuum can be expressed as  an entangled two-mode squeezed state
\begin{eqnarray}\label{inae}
\left|0\right \rangle_K=\frac{1}{\cosh u}\sum_{n=0}^\infty \tanh^n u\left|n\right \rangle_{in}\otimes\left|n\right \rangle_{out},
\end{eqnarray}
where $\left|n\right \rangle_{in}$ and $\left|n\right \rangle_{out}$ are  excited-states for Schwarzschild modes inside and outside the event horizon.
This means that the observers in different coordinates will not agree on the particle content of each of these states.  It is then interesting to investigate to what degree the coherence of the quantum state for continuous variables is changed when the state is described by the observers in different coordinates.

\section{MEASUREMENT OF QUANTUM COHERENCE FOR CONTINUOUS VARIABLES \label{GSteering}}
In this section, we review the measurement of quantum coherence for  continuous variables   \cite{XU}.
A quantum system is called a continuous-variable system because it  has an infinite-dimensional Hilbert space described by observables with continuous eigenspectra \cite{weedbrook}. The prototype of a continuous-variable system is
represented by $N$ bosonic modes corresponding to the field operators
$\{\hat{a}_{k},\hat{a}_{k}^{\dagger}\}_{k=1}^{N}$. These
annihilation and creation operators can be
arranged in a vectorial operator $\mathbf{\hat{b}}:=(\hat{a}_{1},\hat{a}%
_{1}^{\dagger},\cdots,\hat{a}_{N},\hat{a}_{N}^{\dagger})^{T}$, which must
satisfy the bosonic commutation relations
\begin{equation}
\lbrack\hat{b}_{i},\hat{b}_{j}]=\Omega_{ij}, \label{BCR}%
\end{equation}
where $\Omega_{ij}$ are  generic elements of
\begin{equation}
\boldsymbol{\Omega}:=\bigoplus\limits_{k=1}^{N}\boldsymbol \left(
\begin{array}
[c]{cc}%
0 & 1\\
-1 & 0
\end{array}
\right), \label{Symplectic_Form}%
\end{equation}
known as the symplectic form. The Hilbert
space of this system is  infinite-dimensional because
the single-mode Hilbert space $\mathcal{H}$ is spanned by a countable Fock basis
$\{\left\vert n\right\rangle \}_{n=0}^{\infty}$.
Besides the bosonic field operators, the bosonic system may be described by
 the quadrature field operators,  formally arranged in the vector
$
\mathbf{\hat{R}}:=(\hat{q}_{1},\hat{p}_{1},\ldots,\hat{q}_{N},\hat{p}_{N}%
)^{T}$  \cite{weedbrook},
which are related to the annihilation $\hat{a}_{i}$ and creation  $\hat{a}_{i}^{\dag}$
operators of each mode, by the relations $\hat{q}_{i}=\frac{(\hat{a}_{i}+\hat{a}_{i}^{\dag})}{\sqrt{2}}$
and $\hat{p}_{i}=\frac{(\hat{a}_{i}-\hat{a}_{i}^{\dag})}{\sqrt{2}i}$.
The quadrature   operators $\hat{q}_{i}$ and $\hat{p}_{i}$  represent the canonical observables
of the system. Similarly,
the vector operator should  satisfy the commutation relationship
$[{{{\hat R}_i},{{\hat R}_j}} ] = i{\Omega _{ij}} $, which takes  symplectic form. The most relevant quantities that characterize the nature of a two-mode Gaussian state ${\rho _{AB}}$
are the statistical moments. The first moment is the mean value
$
{\bar{R}}:=\langle {\hat{R}}\rangle =\mathrm{Tr}
({\hat{R}}\hat{\rho})$,
and the second moment is the covariance
matrix $\mathbf{\sigma}$, whose arbitrary element is
defined by%
$
\sigma_{ij}=\langle \hat{R}_{i}\hat{R}
_{j}+\hat{R}_{j}\hat{R}
_{i}  \rangle-2\langle \hat{R}_{i} \rangle\langle \hat{R}
_{j} \rangle. \label{CM_definition}
$
The covariance matrix is a
symmetric matrix which must satisfy the uncertainty principle
$
\mathbf{\sigma}+i\mathbf{\Omega}\geq0$  \cite{Simon1994}, which
implies the positive definiteness $\mathbf{\sigma}>0$.

For a two-mode Gaussian state, we can write its covariance matrix in the
block form
\begin{equation}
\boldsymbol{\sigma}\equiv\left(\begin{array}{cc}
\boldsymbol{\alpha}&\boldsymbol{\gamma}\\
\boldsymbol{\gamma}^{T}&\boldsymbol{\beta}
\end{array}\right)\ ,
\end{equation}
where $\mathbf{\alpha}=\mathbf{\alpha}^{T}$, $\mathbf{\beta}=\mathbf{\beta}^{T}$ and
$\mathbf{\gamma}$ are $2\times2$ real matrices. Then, the Williamson form is simply
$\mathbf{\sigma}^{\oplus}=(\nu_{-}\mathbf{I})\oplus(\nu_{+}\mathbf{I})$, where
symplectic spectrum $\{\nu_{-},\nu_{+}\}$ is provided by $
\nu_{\pm}=\sqrt{\frac{\Delta \pm\sqrt{\Delta^{2}%
-4\det\mathbf{V}}}{2}}~, $
with $\Delta:=\det\mathbf{\alpha}+\det\mathbf{\beta}+2\det\mathbf{\gamma}$ and $\det$ is the determinant~\cite{weedbrook}.

As shown in \cite{XU}, the definition of the continuous variable quantum coherence of a Gaussian state is
\begin{equation}
C(\rho)=\inf{S(\rho ||\delta)}, \end{equation} where $S(\rho||\delta)= tr(\rho \log_{2}\rho) - tr(\rho \log_{2}\delta)$ is the relative entropy, $\delta$ is an incoherent Gaussian state  and the minimization runs over all incoherent Gaussian states. In addition, the entropy of $\rho$ is defined by
\cite{entropy}
 \begin{eqnarray}\label{coherence1}
 S(\rho)=-tr(\rho \log_{2}\rho)=\sum_{i=1}^{m}f({\nu}_i),
 \end{eqnarray}
where $f({\nu}_i)=\frac{{\nu}_i+1}{2}\log_2\frac{{\nu}_i+1}{2}-\frac{{\nu}_i-1}{2}\log_2\frac{{\nu}_i-1}{2}$, and $\nu_i$ are symplectic eigenvalues of each modes. The mean occupation value can be expressed as
\begin{eqnarray}
\overline{n}_i=\frac{1}{4}(\sigma_{11}^i+\sigma_{22}^i+[d_1^i]^2+[d_2^i]^2-2).
\end{eqnarray}
In this equation, $\sigma^i$ are elements of the subsystem of mode $i$ in a continuous variable matrix, and $[d^i]^2$ is the $i$-th first statistical moment of the mode.
Then the measurement of  quantum coherence of Gaussian states can be expressed as \cite{XU}
\begin{eqnarray}\label{coherence3}
 C({\rho})=-S(\rho)+\sum_{i=1}^{m}[(\overline{n}_i+1)\log_2(\overline{n}_i+1)
-\overline{n}_i\log_2\overline{n}_i].
\end{eqnarray}

\section{The dynamics  OF QUANTUM COHERENCE IN GHS DILATON BLACK HOLE \label{tools}}
\subsection{The dynamics of quantum coherence between the  modes observed by Alice and Bob}

In this subsection, we seek for  a phase-space description for the dynamics of Gaussian quantum  coherence under the influence of  the  GHS dilaton black hole. We assume that an  observer Alice  stays  at the asymptotically flat region, while  Bob  observing subsystem $B$ hovers near the event horizon of the black hole.
The  initial state shared between them is a two-mode squeezed Gaussian state
\begin{eqnarray}\label{inAR}
\sigma^{\rm (G)}_{AB}(s)=
\left( {\begin{array}{*{20}{c}}
 I_{2} \cosh 2s & Z_2 \sinh 2s    \\
Z_2  \sinh 2s  & I_{2}  \cosh 2s\\
\end{array}} \right),
\end{eqnarray}
 where
$Z_2 =\left(
                       \begin{array}{cc}
                         1 & 0 \\
                         0 & -1 \\
                       \end{array}
                     \right)$,  
$I_2 =\left(
                       \begin{array}{cc}
                         1 & 0 \\
                         0 & 1 \\
                       \end{array}
                     \right)$
 and  $s$ is the squeezing parameter.

It has been shown in  Eq. (\ref{inae}) that  the Kruskal vacuum is an entangled two-mode squeezed state in terms of Schwarzschild modes. The two mode  squeezed transformation can be expressed by a symplectic operator in the phase-space
\begin{eqnarray}\label{cmtwomode}
S_{B,\bar B}(D)=\left(\!\!\begin{array}{cccc}
I_{2} \cosh u   &Z_{2} \sinh u\\
Z_{2}\sinh u& I_{2}  \cosh u
\end{array}\!\!\right).\end{eqnarray}

After the action of the  two-mode  squeezed transformation,  the entire system involves three subsystems: the subsystem $A$ described by the global observer Alice, the subsystem $B$ described by Bob, and the subsystem $\bar B$  described by the virtual observer anti-Bob. The covariance matrix $\sigma_{AB\bar B}$ of the tripartite quantum system is given by \cite{adesso3}
\begin{eqnarray}\label{in34}
\sigma_{AB \bar B}(s,D) &=& \big[I_A \oplus  S_{B,\bar B}(D)\big] \big[\sigma^{\rm (G)}_{AB}(s) \oplus I_{\bar B}\big]\\&& \nonumber\big[I_A \oplus  S_{B,\bar B}(D)\big],
\end{eqnarray}
where $S_{B,\bar B}(D)$ is the phase-space representation of the two-mode squeezing operation given in Eq. (\ref{cmtwomode}). In Eq. (\ref{in34}), the matrix $\big[\sigma^{\rm (G)}_{AB}(s) \oplus I_{\bar B}\big]$  describes the initial state for the entire system, and $ \big[I_A \oplus  S_{B,\bar B}(D)\big] $ denotes that the two-mode  squeezed transformation is  only performed on  the  bipartite subsystem between Bob and anti-Bob.

Because the exterior region of the black hole is causally disconnected  to the inner region, Alice and Bob cannot approach the mode $\bar B$. Then one obtains the covariance matrix $\sigma_{AB}(s,D)$ for Alice and Bob by tracing out the mode $\bar B$
\begin{equation}\label{CM1}
\sigma_{AB}(s,D)= \left( {\begin{array}{*{20}{c}}
   \mathcal{A}_{AB} & \mathcal{C}_{AB}  \\
   {{\mathcal{C}_{AB}^{\sf T}}} & \mathcal{B}_{AB} \\
\end{array}} \right),
\end{equation} where  \begin{equation} \nonumber \mathcal{A}_{AB}=\cosh2sI_2,\end{equation}
\begin{equation} \nonumber \mathcal{C}_{AB}=\cosh u\sinh 2s Z_2, \end{equation}
and \begin{equation} \nonumber \mathcal{B}_{AB}=[{\cosh^2 u\cosh2s+\sinh^2 u}]I_2.\end{equation}

We know that the symplectic matrix $\sigma_{AB}(s,D)$ is a case of the smallest mixed Gaussian state according to the partially transposed symplectic matrix. From the formula above, we
obtain
\begin{eqnarray}\label{CC}
   \Delta^{(AB)}=1+[\cosh^2 u+\sinh^2 u\cosh2s]^2.
\end{eqnarray}

The mean occupation numbers operator for each mode from the covariance matrix are
$\bar{n}_{A}=\sinh^2s$ and
$ \bar{n}_{B}=\cosh^2 u\cosh^2s-1$.
Inserting them into Eq. (\ref{coherence3}), we obtain the quantum coherence of the Gaussian state Eq. (31).
Then we can see that the  mean occupation numbers as well as the  Gaussian coherence depend not only on the dilaton charge $D$ and the mass $M$ of the black hole but also on the frequency $\omega$ and the squeezing parameter $s$. 

\begin{figure}[htbp]
\center
\includegraphics[height=2.1in,width=2.8in]{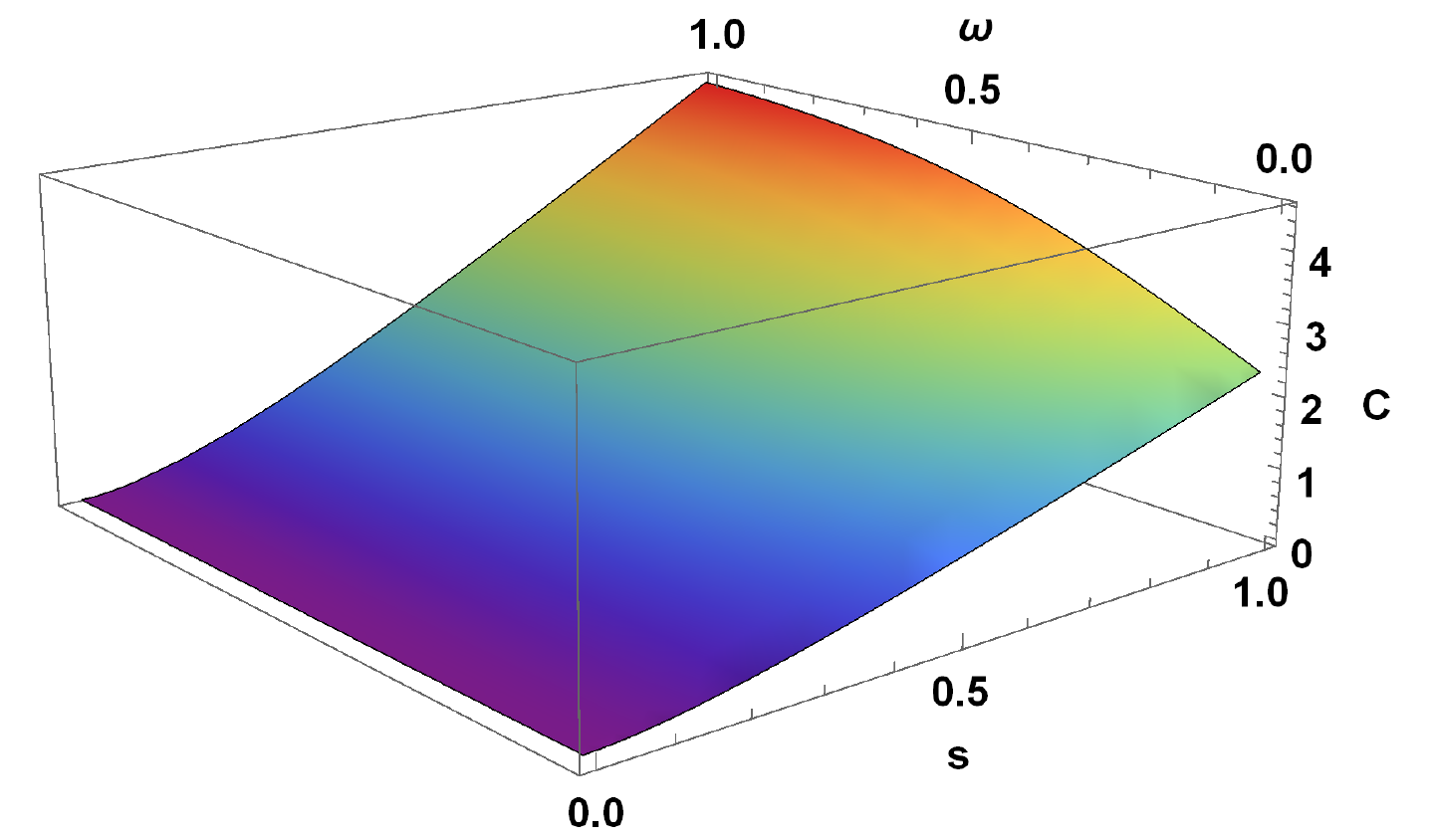}
\caption{ The Gaussian quantum  coherence between Alice and Bob for the different $\omega$ and $s$.  The GHS dilaton charge parameter and mass parameter are fixed as $D/M=0.8$.}\label{Fig1}
\end{figure}

In Fig. (1), we plot the accessible quantum coherence between Alice and  Bob as a function of the initial squeezing parameter $s$ and the frequency $\omega$ of the scalar field. It is illustrated that the  Gaussian  quantum  coherence is a monotonic increasing function of the frequency. This indicates that one can obtain more coherence by choosing a field with a higher frequency. It is worth noting that the coherence is almost  unaffected  to the change of frequency  if the squeezing parameter is very small.
  On the other hand, it becomes more sensitive to the change of the frequency  when the squeezing parameter becomes larger. That is, the frequency of the field produces  positive effects on the storage of quantum coherence in the GHS black hole.

\begin{figure}[htbp]
\centering
\includegraphics[height=2.2in,width=2.8in]{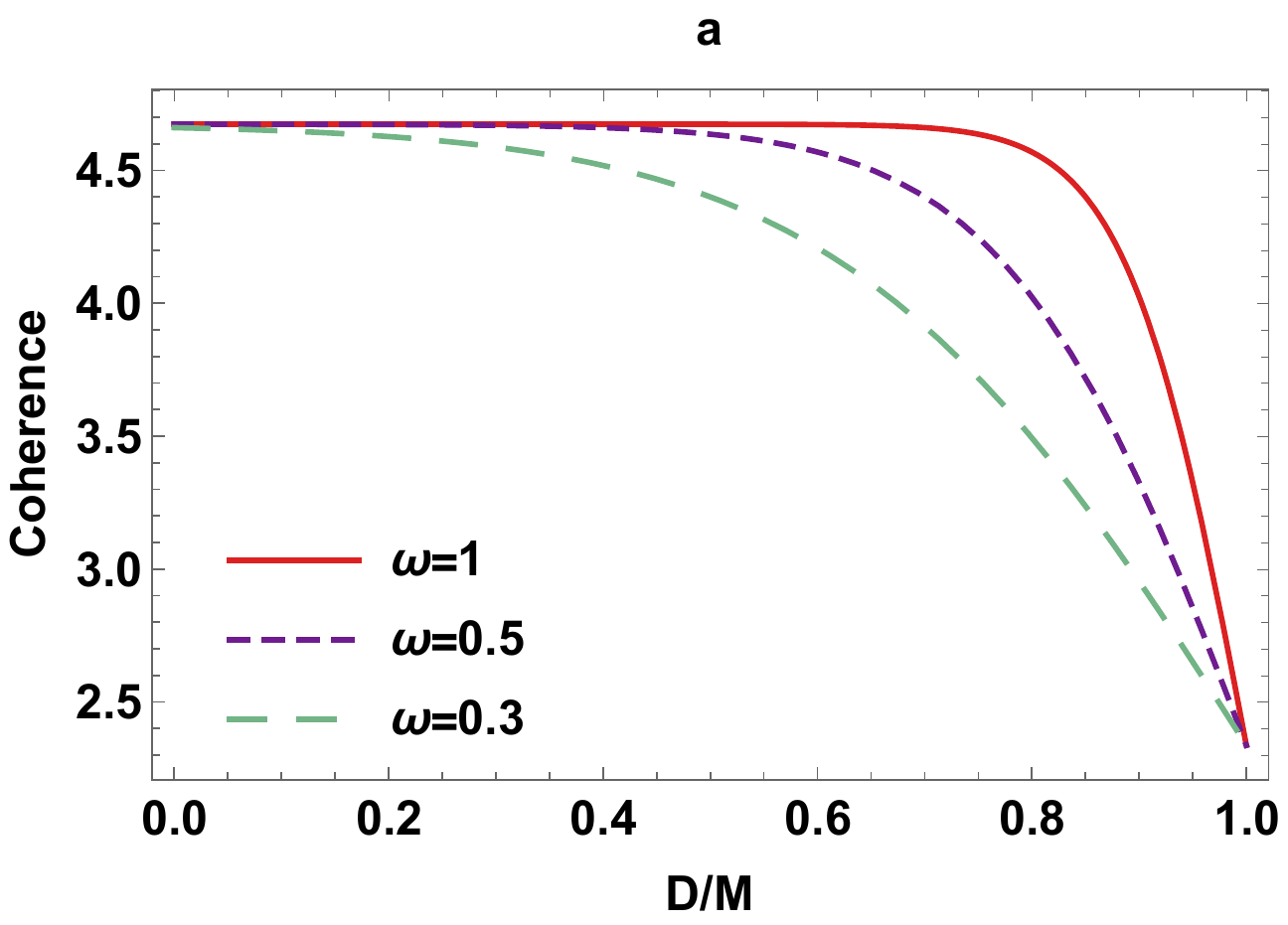}
\includegraphics[height=2.1in,width=2.6in]{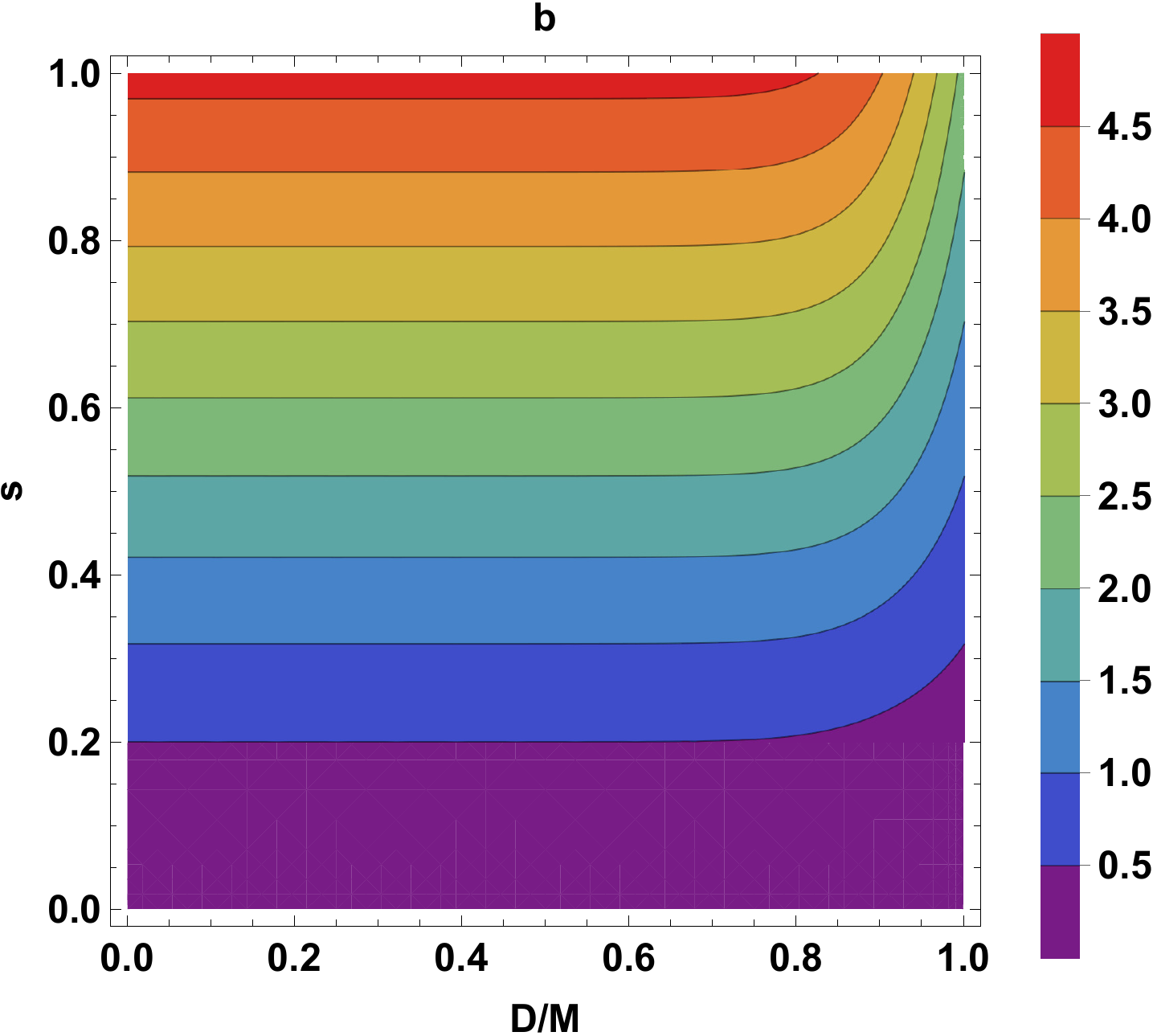}
\caption{ (Color online). (a) Plots of the quantum  coherence between Alice and Bob as a function of the ratio $D/M$ between the dilaton charge and the mass of the black hole. The initial squeezing parameter is fixed as $s=1$ and the frequency of the field are $\omega=1$ (red line), $\omega=0.5$ (purple dotted line), $\omega=0.3$ (green dotted line), respectively. (b) The contour diagram of  Gaussian quantum coherence between Alice and Bob versus the squeezing parameter $s$ and the ratio $D/M$. The frequency of the scalar field is fixed as $\omega=1$.}\label{Fig2}
\end{figure}

In Fig. 2(a), we plot the quantum coherence between the initial correlated  Alice and Bob as a function of  the ratio between the dilaton charge and the mass of the black hole. We find that Gaussian  quantum coherence in the state $\rho_{AB}$  is not affected by the frequency $\omega$ when the ratio is $D/M=0$. In addition, the coherence is independent of the frequency $\omega$ in the limit of  $D\to M$, which corresponds to an extreme dilaton black hole. That is to say, the coherence is not affected by the frequency of the scalar field for an uncharged  black hole or an extreme dilaton black hole.

It is found that the Gaussian  coherence between Alice and Bob decreases with the growth of the ratio $D/M$,
which means that  the gravitational field induced by  dilaton will destroy the quantum coherence  between the initially correlated  modes. This verifies the fact that the gravitational field induced by dilaton plays a key role in the dynamics of Gaussian coherence in the dilaton spacetime. It is interesting to note that the Gaussian  coherence between Alice and Bob is not completely destroyed even in the  limit of  $D\to M$,  which corresponds to an extreme black hole. This is quite different from the behavior of quantum steering in the GHS spacetime because the latter suffers from a ``sudden death" \cite{biwei}. This is  also different from the behavior of quantum entanglement in the GHS spacetime because the entanglement decays to zero only in the limit of $D\to M$ \cite{wang19}.  That is to say, quantum coherence  is more robust than entanglement and steering  under the influence of spacetime effects near the event horizon of the dilaton black hole.

This is reasonable because quantum resources are hierarchic.  It is known that quantum coherence can be defined for the integral system, while quantum correlations characterize the quantum features of a bipartite or a multipartite system \cite{coherencer1}. The results in the present paper, as well as those in Refs. \cite{wang19,biwei}, verify the fact that the quantum resources are hierarchic and quantum coherence is more robust than quantum correlations in the dilaton spacetime. On the other hand, this result indicates the possibility of Bob being capable of performing quantum information tasks even in the case of an extreme dilaton black hole because quantum coherence is a usable resource for the tasks. This is nontrivial because one can employ quantum coherence as the resource of quantum information processing tasks even if the quantum correlations have been destroyed by strong gravitational effects.

Fig. 2(b) shows a contour diagram of  Gaussian coherence versus the squeezing parameter $s$ and the ratio $D/M$.
We can see that, if the initial squeezing parameter is close to 1, the gravitational field of the spacetime has significant influence on the Gaussian quantum coherence only near the   $D\rightarrow M$ limit. The coherence between Alice and Bob is nonzero even in the limit of an extreme dilaton black hole.  This means that one can perform quantum information processing tasks in the GHS spacetime if sufficient resource is prepared in the initial state.  In addition, the coherence becomes more sensitive with the change of the squeezing parameter $s$ for larger dilaton parameters.

\subsection{Generating quantum coherence between the initially uncorrelated modes }

In this subsection, we study the dynamics of quantum coherence among the initially uncorrelated modes. The covariance matrix between Alice and the observer anti-Bob  is obtained by tracing over the outside mode $B$
\begin{equation}\label{CM22}
\sigma_{A\bar B}(s,r) = \left( {\begin{array}{*{20}{c}}
   \mathcal{A}_{A\bar B} & \mathcal{C}_{A\bar B}  \\
   {{\mathcal{C}_{A\bar B}^{\sf T}}} & \mathcal{B}_{A\bar B}  \\
\end{array}} \right),
\end{equation} where
\begin{equation} \nonumber \mathcal{A}_{A\bar B}=[\cosh(2s)]I_2,\end{equation}  \begin{equation} \nonumber \mathcal{C}_{A\bar B}=[\sinh u\sinh2s]I_2,\end{equation}
and \begin{equation} \nonumber \mathcal{B}_{A\bar B}=[\sinh^2 u\cosh2s+\cosh^2 u]I_2.\end{equation}

The corresponding symplectic value of the covariance matrix $\sigma_{A\bar B}$ is found to be  $\nu_{+}={\cosh^2 u\cosh2s+\sinh^2 u}$ and $\nu_{-}=1$.
Similarly, we find
$
\Delta^{(A\bar{B})}=1+[\cosh^2 u\cosh2s+\sinh^2 u]^2$. Then we calculate the quantum coherence between the Alice and anti-Bob and plot it in Fig (3).

\begin{figure}[htbp]
\centering
\includegraphics[height=2.2in,width=2.8in]{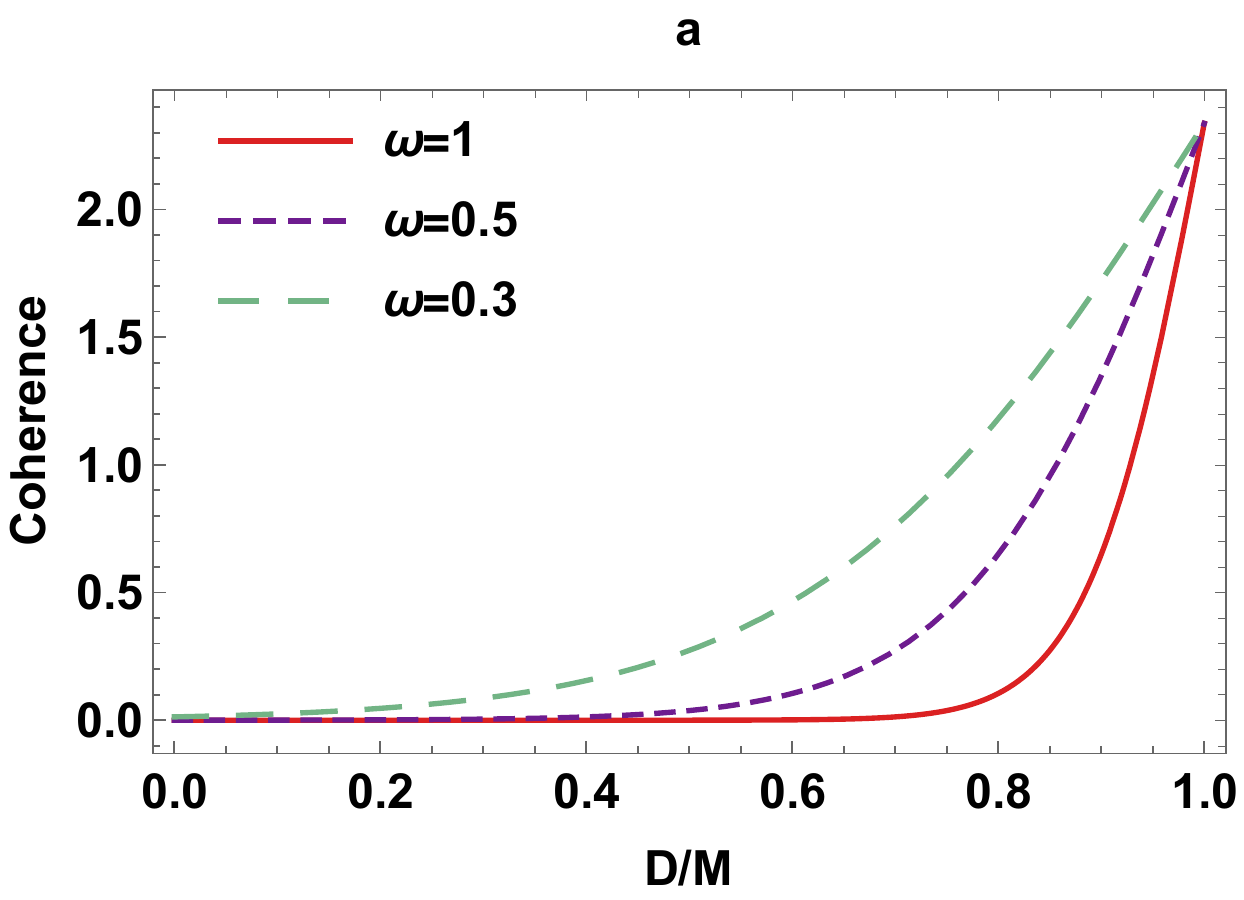}
\includegraphics[height=2.1in,width=2.6in]{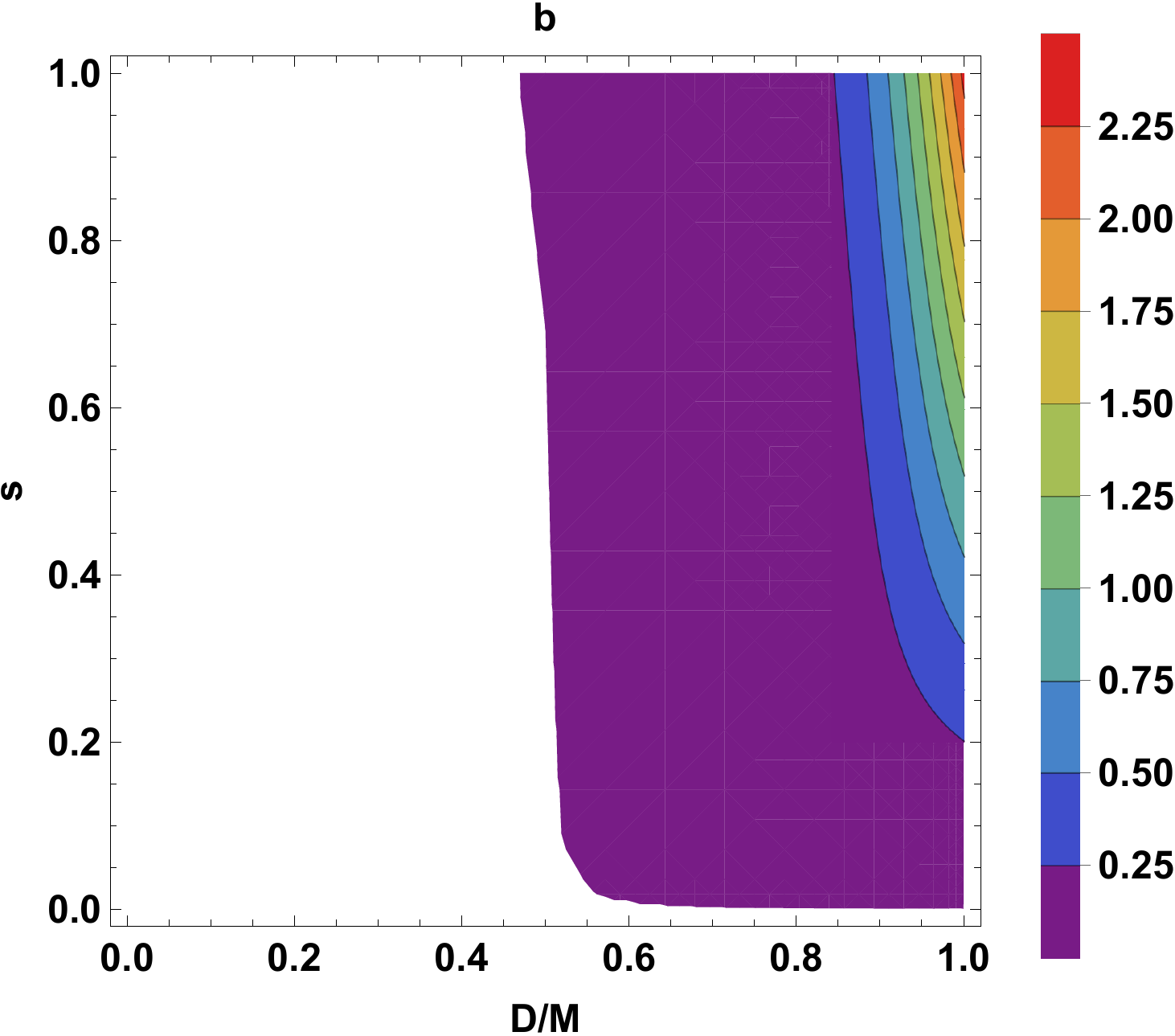}
\caption{ (Color online). (a) Plots of the quantum  coherence between Alice and anti-Bob as a function of the ratio $D/M$. The initial squeezing parameter is fixed as $s=1$ and the frequency of the field are $\omega=1$ (red line), $\omega=0.5$ (purple dotted line), $\omega=0.3$ (green dotted line), respectively. (b) The contour diagram of  Gaussian quantum coherence between Alice and anti-Bob versus the squeezing parameter $s$ and the ratio $D/M$.  The frequency of the scalar field is fixed as $\omega=1$. }\label{Fig3}
\end{figure}

Fig. 3(a) shows the Gaussian  quantum coherence between Alice and anti-Bob as a function of the  ratio $D/M$ for different frequencies. Again,  it is found that the quantum coherence is independent of the frequency $\omega$ for the non-dilaton and extreme dilaton black hole.  As the increase of the  ratio $D/M$, the Gaussian coherence is generated between Alice and anti-Bob, which is quite different from the quantum steering between Alice and Bob for the same state \cite{biwei}, where the Gaussian  quantum  steering is zero for any dilaton charge.
In addition,  the lower the frequency of the mode in the state $\rho_{A\bar B}$, the stronger the Gaussian quantum coherence. This indicates the generation of quantum coherence between Alice and Bob is easier for low-frequency fields. In Fig. 3(b),  it is shown that the Gaussian coherence between Alice and anti-Bob  exhibits a ``sudden birth" behavior under the influence of the dilaton field.

The covariance matrix between Bob and anti-Bob inside the event horizon is obtained by tracing over the mode $A$
\begin{equation}\label{CM22}
\sigma_{B\bar B}(s,r) = \left( {\begin{array}{*{20}{c}}
   \mathcal{A}_{B\bar B} & \mathcal{C}_{B\bar B}  \\
   {{\mathcal{C}_{B\bar B}^{\sf T}}} & \mathcal{B}_{B\bar B}  \\
\end{array}} \right),
\end{equation} where
\begin{equation} \nonumber \mathcal{A}_{B\bar B}=[\cosh^2 u\cosh2s+\sinh^2 u]I_2,\end{equation}
\begin{equation} \nonumber \mathcal{C}_{B\bar B}=[\sinh 2u\cosh^2s]Z_2,\end{equation}
and \begin{equation} \nonumber \mathcal{B}_{B\bar B}=[\sinh^2 u\cosh2s+\cosh^2 u]I_2.\end{equation}
We can also calculate the Gaussian quantum coherence between Bob and anti-Bob.

\begin{figure}[htbp]
\centering
\includegraphics[height=2.1in,width=2.8in]{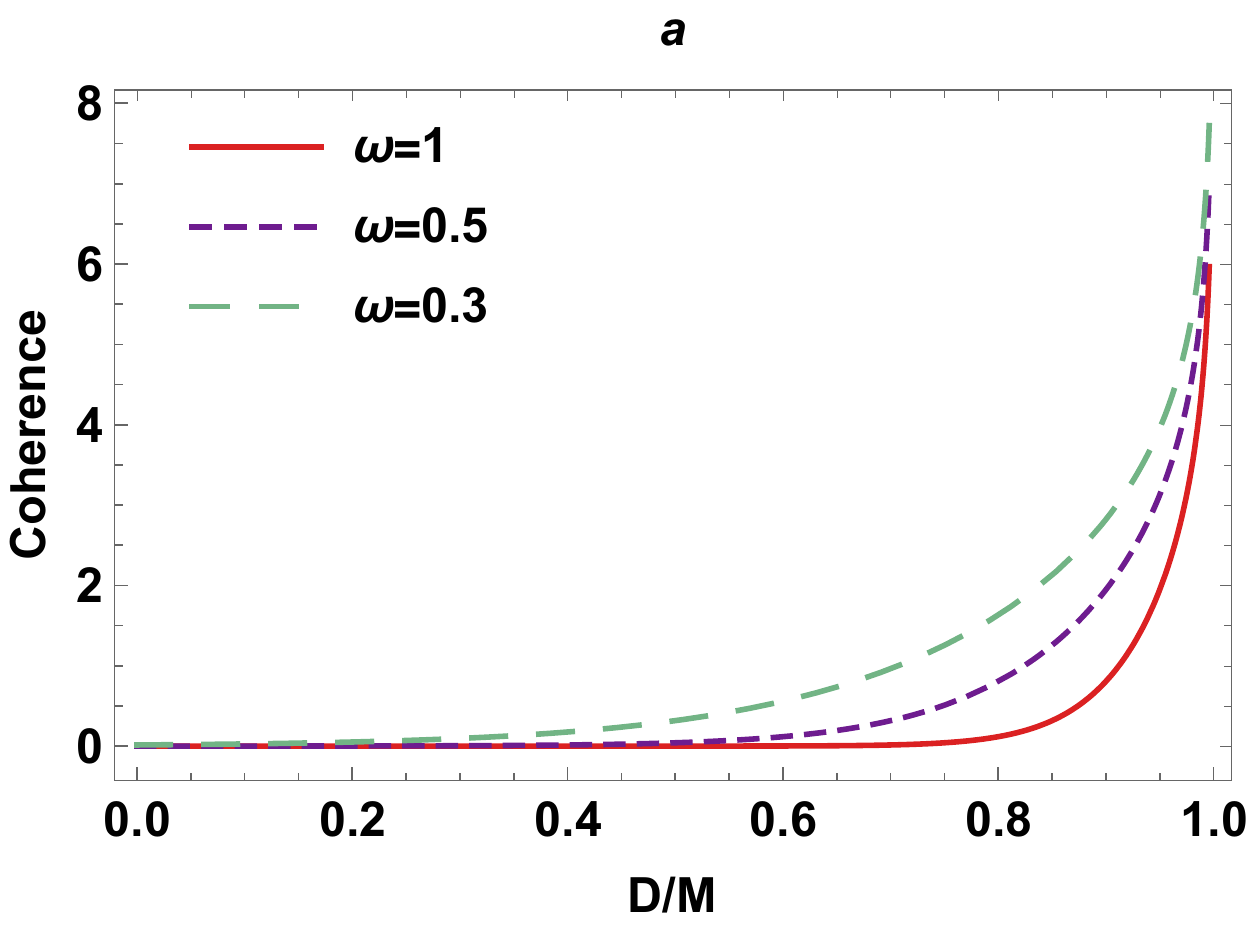}
\includegraphics[height=2.1in,width=2.6in]{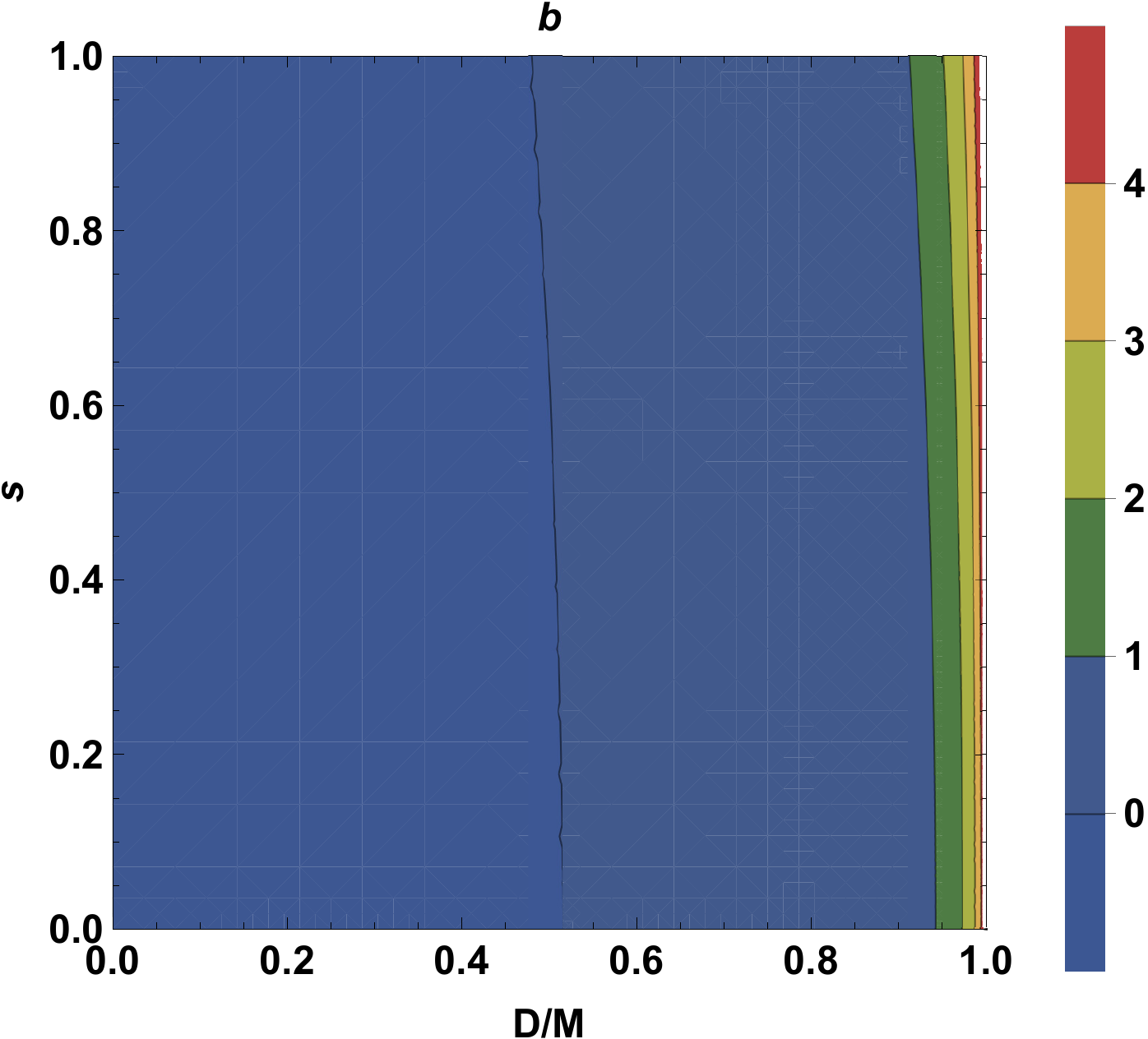}
\caption{ (Color online). (a) Plots of the quantum  coherence between Bob and anti-Bob as a function of the ratio $D/M$. The initial squeezing parameter is fixed as $s=1$. (b) The contour diagram of  Gaussian quantum coherence between Bob and anti-Bob versus the squeezing parameter $s$ and the ratio $D/M$.  The frequency of the scalar field is fixed as $\omega=1$. }\label{Fig3}
\end{figure}

In Fig. 4, we show that the quantum coherence between  Bob and anti-Bob is a monotonically increasing function of $D/M$.  It changes slowly at first and becomes more sensitive with the increase of $D/M$. It is also found that the quantum coherence is independent of the frequency $\omega$ for the non-dilaton and extreme dilaton black hole.  As the increase of the ratio $D/M$, the quantum  coherence is smoothly  generated between Bob and anti-Bob. This is different from the  generation of  Gaussian coherence between Alice and anti-Bob because the latter exhibits a ``sudden birth" behavior.  It is shown that the gravitational field induced by dilaton generates Gaussian quantum coherence between the causally disconnected regions. In other words, Bob and anti-Bob can perform quantum information processing tasks  by local measurements even though they are separated by the event horizon.

\section{SUMMARY}
In this work, we study  the behavior of quantum coherence for Gaussian states in the background of a GHS dilaton black hole.  It is shown that the dilaton field has evident effect on the degree of coherence for all the bipartite subsystems. However, the Gaussian coherence is not affected by the frequency of the scalar field for an uncharged or an extreme dilaton black hole. This verifies the fact that the gravity induced by dilaton field plays a key role in the dynamics of Gaussian coherence in the GHS spacetime. The Gaussian  coherence between Alice and Bob is not completely destroyed even for an extreme dilaton black hole. This is quite different from the behavior of quantum steering because the latter suffers from a ``sudden death"  under the same condition. The coherence between Alice and Bob is nonzero even in the limit of an extreme dilaton black hole. This means that one can perform quantum information processing tasks in the GHS spacetime if sufficient resources are prepared in the initial state. 

\begin{acknowledgments}
This work is supported by   the National Natural Science Foundation
of China under Grant  No. 12122504, No. 12035005,  and No. 11875025.	
\end{acknowledgments}

\end{document}